\documentclass[aps,prb,twocolumn,superscriptaddress,showpacs]{revtex4-1}
\usepackage{graphicx,bm,mathtools,pbox}

\begin{document}

\title{Evidence for nodal superconductivity in quasi-one-dimensional K$_2$Cr$_3$As$_3$}

\author{G. M. Pang}
\affiliation{Center for Correlated Matter and Department of Physics, Zhejiang University, Hangzhou, 310058, China}
\affiliation{Collaborative Innovation Center of Advanced Microstructures, Nanjing 210093, China}
\author{M. Smidman}
\affiliation{Center for Correlated Matter and Department of Physics, Zhejiang University, Hangzhou, 310058, China}
\affiliation{Collaborative Innovation Center of Advanced Microstructures, Nanjing 210093, China}
\author{W. B. Jiang}
\affiliation{Center for Correlated Matter and Department of Physics, Zhejiang University, Hangzhou, 310058, China}
\affiliation{Collaborative Innovation Center of Advanced Microstructures, Nanjing 210093, China}
\author{J. K. Bao}
\affiliation{Center for Correlated Matter and Department of Physics, Zhejiang University, Hangzhou, 310058, China}
\affiliation{Collaborative Innovation Center of Advanced Microstructures, Nanjing 210093, China}
\author{Z. F.  Weng}
\affiliation{Center for Correlated Matter and Department of Physics, Zhejiang University, Hangzhou, 310058, China}
\affiliation{Collaborative Innovation Center of Advanced Microstructures, Nanjing 210093, China}
\author{Y. F. Wang}
\affiliation{Center for Correlated Matter and Department of Physics, Zhejiang University, Hangzhou, 310058, China}
\affiliation{Collaborative Innovation Center of Advanced Microstructures, Nanjing 210093, China}
\author{L. Jiao}
\affiliation{Center for Correlated Matter and Department of Physics, Zhejiang University, Hangzhou, 310058, China}
\affiliation{Collaborative Innovation Center of Advanced Microstructures, Nanjing 210093, China}
\author{J. L. Zhang}
\affiliation{Center for Correlated Matter and Department of Physics, Zhejiang University, Hangzhou, 310058, China}
\affiliation{Collaborative Innovation Center of Advanced Microstructures, Nanjing 210093, China}
\author{G. H. Cao}
\affiliation{Center for Correlated Matter and Department of Physics, Zhejiang University, Hangzhou, 310058, China}
\affiliation{Collaborative Innovation Center of Advanced Microstructures, Nanjing 210093, China}
\author{H. Q. Yuan}
\email{hqyuan@zju.edu.cn}
\affiliation{Center for Correlated Matter and Department of Physics, Zhejiang University, Hangzhou, 310058, China}
\affiliation{Collaborative Innovation Center of Advanced Microstructures, Nanjing 210093, China}

\date{\today}

\begin{abstract}
The recent discovery of superconductivity in the quasi-one-dimensional compound K$_2$Cr$_3$As$_3$, which consists of double-walled tubes of [(Cr$_3$As$_3$)$^{2-}]^\infty$ that run along the c axis, has attracted immediate attention as a potential system for studying superconductors with reduced dimensionality.  Here we report clear experimental evidence for the unconventional nature of the superconducting order parameter in K$_2$Cr$_3$As$_3$, by precisely measuring the temperature dependence of the change in the penetration depth $\Delta\lambda(T)$ using a tunnel diode oscillator. Linear behavior of $\Delta\lambda(T)$ is observed for $T\ll T_c$, instead of the exponential behavior of conventional superconductors, indicating that there are line nodes in the superconducting gap. This is strong evidence for unconventional behavior and may provide key information for identifying the pairing state of this novel superconductor.
\end{abstract}


\maketitle

Superconductivity in low-dimensional systems has been the focus of intense experimental and theoretical research and often displays unconventional properties. Both the cuprate and iron-pnictide high temperature superconductors are layered structures and in the case of the former, conduction mostly occurs within planes of CuO$_2$. \citep{Cupr2,IronSC} Sr$_2$RuO$_4$ is another quasi-two-dimensional material which is believed to display triplet superconductivity, \citep{Sr2RuO41,Sr2RuO42} whereas the $T_c$ of heavy fermion superconductors appears to be enhanced when reducing the dimensionality. \citep{NatureMagSC,CeCoIn5Ref} There are fewer quasi-one-dimensional (q1D) systems reported to display superconductivity. Some examples include Li$_{0.9}$Mo$_6$O$_{17}$\citep{LiMoORep} and the organic superconductors (TMTSF)$_2X$ (TMTSF = tetramethyltetraselenafulvalene, $X$ = PF$_6$, ClO$_4$). \citep{OrgRef1,OrgRef2} Evidence for spin triplet superconductivity in both these systems comes from the highly anisotropic upper critical fields \citep{Hc2Org,Hc2LiMoO} which vastly exceed the Pauli limiting field along certain crystallographic directions. However, the pairing symmetry of (TMTSF)$_2X$ remains unresolved and for $X~=~$ClO$_4$, there is evidence for spin singlet superconductivity with nodes in the superconducting gap. \citep{TMTSFClO4Sing1,TMTSFClO4Sing2,TMTSFClO4Node}

\begin{figure}[h]
\begin{center}
  \includegraphics[width=0.75\columnwidth]{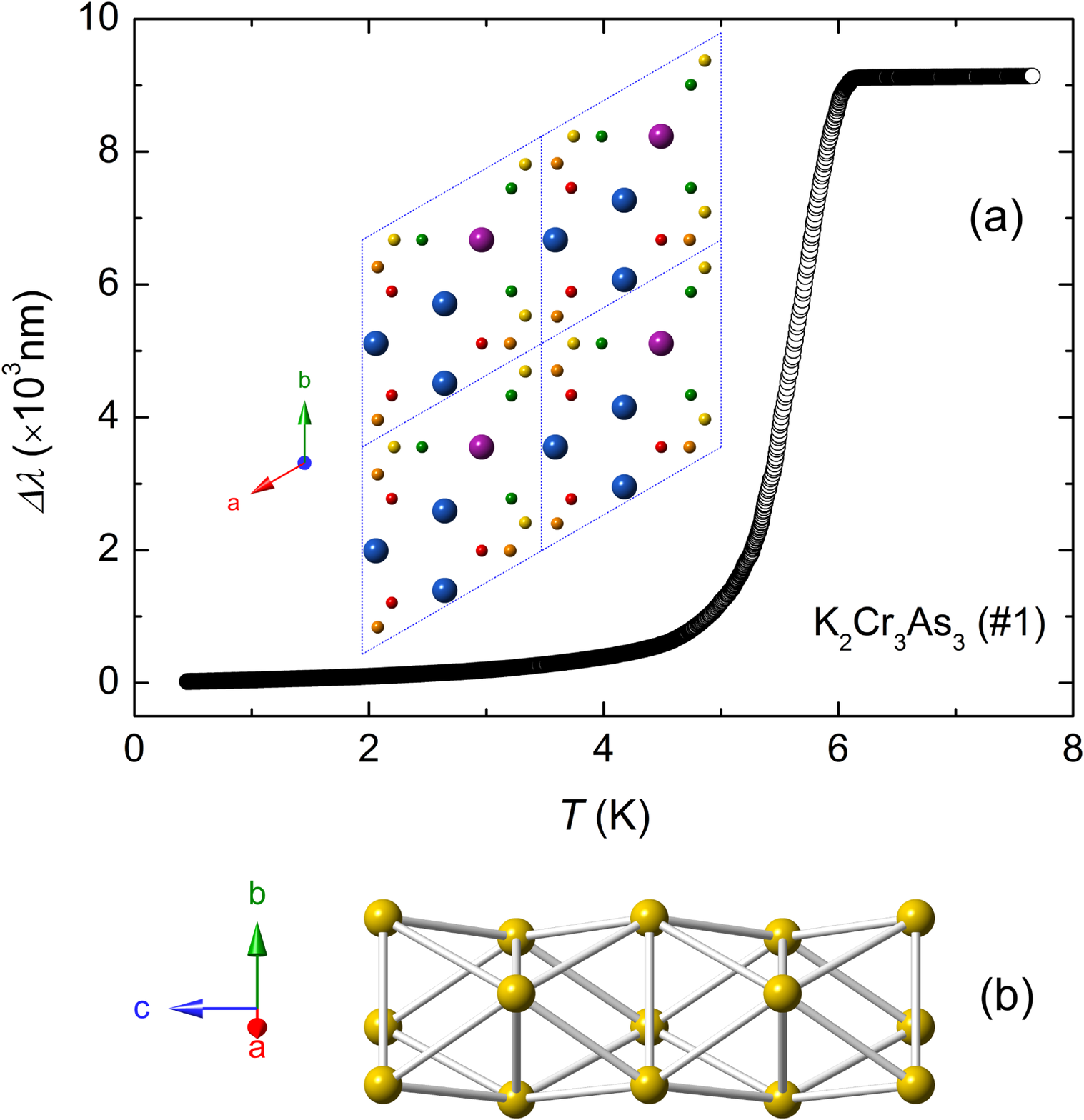}
\end{center}
	\caption{(a)Temperature dependence of the London penetration depth $\Delta\lambda(T)=\lambda(T)-\lambda(0)$  across the superconducting transition  for polycrystalline K$_2$Cr$_3$As$_3$(sample~\#1). The inset shows the crystal structure of K$_2$Cr$_3$As$_3$ as viewed along the $c$~axis. For the $z~=~0$ sites, the K atoms are in purple, the Cr in yellow and the As in red, while for the $z~=~0.5$ sites, the K are in blue, the Cr in orange and the As in green. The double-walled subnano-tubes of Cr and As atoms can be seen at the intersection of four unit cells. These tubes are separated by the positively charged K$^+$ cations. (b) The arrangement of a Cr tube, as viewed perpendicular to the $c$~axis. The Cr atoms form face sharing octahedra.}
   \label{StrucFig}
\end{figure}

Recently, a class of CrAs based superconductors have attracted considerable attention. CrAs itself is a three-dimensional antiferromagnet which becomes superconducting when pressure is applied to suppress the antiferromagnetic order, with $T_c$ reaching 2~K at a critical pressure of 0.8~GPa.\citep{CrAsSC} Subsequently, bulk superconductivity was also discovered in the related hexagonal compounds $A_2$Cr$_3$As$_3$ ($A$~=~K, Rb), with $T_c=6.1$ ~K for K$_2$Cr$_3$As$_3$ and 4.8~K for Rb$_2$Cr$_3$As$_3$. \citep{K2Cr3As3Rep, Rb2Cr3As3Rep} The crystal structure of  $A_2$Cr$_3$As$_3$ consists of [(Cr$_3$As$_3$)$^{2-}]^\infty$ double-walled subnano-tubes separated by K$^+$ or Rb$^+$ cations. These tubes have a concentric arrangement with the outside tubes consisting of As and Cr tubes on the inside, as shown in Fig.~\ref{StrucFig}. These structural features suggest that  $A_2$Cr$_3$As$_3$ may be regarded as q1D systems. The most remarkable of the reported superconducting properties is the large value of the upper critical field at zero temperature [$H_{c2}(0)]$. For example, $T_c$ of K$_2$Cr$_3$As$_3$ is not suppressed much below 4~K in an applied field of 14~T, \citep{K2Cr3As3Crys} despite a BCS Pauli limiting field of $1.84T_c\sim11.2$~T. This gives an indication that Pauli limiting may be absent, which would be evidence for triplet superconductivity. The values of $H_{c2}$ were only found to be weakly anisotropic, which is unlike the very large anisotropy observed in the q1D superconductors (TMTSF)$_2$PF$_6$ and Li$_{0.9}$Mo$_6$O$_{17}$. \citep{Hc2Org,Hc2LiMoO}  Electronic structure calculations for K$_2$Cr$_3$As$_3$  reveal that three bands cross the Fermi level ($E_F$), two of which are one-dimensional and one is three-dimensional.  \citep{K2Cr3As3Elec} Identifying which of these bands are involved in the superconducting pairing would allow for the determination of whether $A_2$Cr$_3$As$_3$ can be regarded as q1D superconductors.

An additional feature of $A_2$Cr$_3$As$_3$ is the presence of enhanced electronic correlations. For example, K$_2$Cr$_3$As$_3$ has a large electronic specific heat coefficient $\gamma$ of about 70~mJ/mol K$^2$. \citep{K2Cr3As3Rep,K2Cr3As3Crys} This is in contrast to Li$_{0.9}$Mo$_6$O$_{17}$ ($\gamma$~=~6~mJ/mol K$^2$) \citep{LiMoOspecH} which is a weakly correlated, q1D material and raises the prospect that $A_2$Cr$_3$As$_3$ may be correlated q1D superconductors. Furthermore, electronic structure calculations suggest that K$_2$Cr$_3$As$_3$ is located close to a ferromagnetic instability.  \citep{K2Cr3As3Elec} This indicates that the superconductivity of these compounds may be mediated not by the conventional electron-phonon pairing mechanism but by spin fluctuations. \citep{Theor2} In addition, the crystal structure of $A_2$Cr$_3$As$_3$ lacks an inversion center (space group $P\bar{6}m2$), which along with a finite antisymmetric spin-orbit coupling (ASOC) can lead to pairing states with a mixture of spin singlet and triplet components. \citep{NCS2001} This may give rise to unconventional behavior such as nodes in the superconducting gap, large anisotropic upper critical fields and an enhanced spin susceptibility below $T_c$. \citep{BauerNCS}

The presence of both a quasi-one-dimensional crystal structure without an inversion center and enhanced electronic correlations in $A_2$Cr$_3$As$_3$ is a rare combination, which may allow for the exploration of novel superconducting properties. Of particular importance in determining the nature of the superconducting state is to gain knowledge of the structure of the superconducting gap. Here we probe the gap symmetry of these q1D compounds by measuring the precise temperature dependence of the London penetration depth of polycrystalline K$_2$Cr$_3$As$_3$ using a tunnel-diode based self-inductive technique.

\begin{figure}[t]
\begin{center}
 \includegraphics[width=0.8\columnwidth]{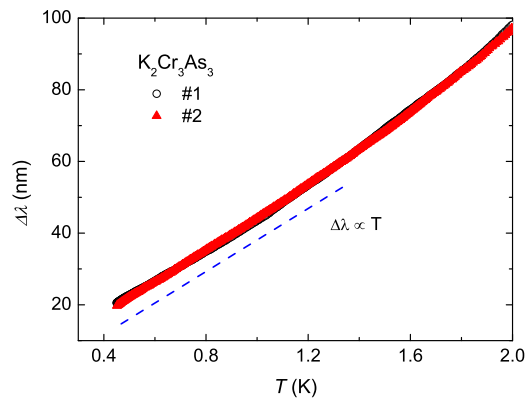}
\end{center}
\caption{Temperature dependence of the London penetration depth change $\Delta\lambda(T)$ for K$_2$Cr$_3$As$_3$ (\#1 and \#2) from 0.4~-~2~K. Linear behaviour is observed up to about 1.4~K, which is evidence for line nodes in the superconducting gap. This is shown by the dashed line, which has been offset from the data.}
\label{LamT}
\end{figure}

Polycrystalline samples were synthesized by a solid state reaction, as described in Ref.~\onlinecite{K2Cr3As3Rep}. The grains are needle-like with a typical length of around $40-50~\mu$m and a thickness of a few $\mu$m, obtained from scanning electron microscopy. \cite{Supp1} The samples are extremely air sensitive, and visibly decompose within seconds of exposure to the air. These were carefully handled in an argon glove box and were encased in Apiezon N~grease when being measured, which does not contribute significantly to the penetration depth measurements. \cite{Supp1}  Measurements of the penetration depth change $\Delta\lambda(T)$ were performed down to 0.4~K in a $^3$He cryostat, using a tunnel-diode based self-inductive technique \citep{TDOMethod}. The sample was mounted on a sapphire rod in order to be inserted into the coil without making contact. The operating frequency of the tunnel diode oscillator was about 7~MHz, with a noise level as low as 0.1~Hz with steady control of the circuit and coil temperature. A small alternating magnetic field of about 20~mOe is applied to the sample, which is much smaller than the lower critical field, ensuring that the material remains in the Meissner state and therefore that $\lambda(T)$ can be taken to be the London penetration depth. The change in $\lambda(T)$ is given by $\Delta\lambda(T)=G\Delta f(T)$, where the calibration constant $G$ depends solely on the sample and coil geometry and was calculated by approximating the sample as a square plate. \citep{Gfactor}

The temperature dependence of $\Delta\lambda(T)$ is displayed in Fig.~\ref{StrucFig} from 0.4 ~K  to 8~K for K$_2$Cr$_3$As$_3$. The decrease indicates the onset of superconductivity at 6.1~K, in agreement with previously reported measurements \citep{K2Cr3As3Rep} and the sharpness of the transition indicates a high sample quality. The midpoint of the transition is about 5.6~K and this is taken as the value of $T_c$ for subsequent calculations. The low temperature behavior of $\Delta\lambda(T)$ is displayed in Fig.~\ref{LamT} for two samples, labeled \#1 and \#2. Values of $G=7.2$~\AA/Hz and $4.0$~\AA/Hz were determined for samples $\#1$ and $\#2$, respectively. It is obvious that in both samples the penetration depth $\Delta\lambda(T)$ shows a linear temperature dependence from the base temperature up to about 1.4~K, as opposed to the exponential temperature dependence observed in BCS superconductors. Non-exponential behavior at low temperatures indicates the presence of low lying electronic excitations and in particular, $\Delta\lambda(T)\sim T$ provides evidence for line nodes in the superconducting gap, although inter-grain Josephson coupling in a polycrystalline sample might also give rise to a linear contribution. \citep{JosGran} We note that the linear behavior at low temperatures is reproducible in high quality samples which display a sharp superconducting transition.

\begin{figure}[t]
\begin{center}
 \includegraphics[width=0.8\columnwidth]{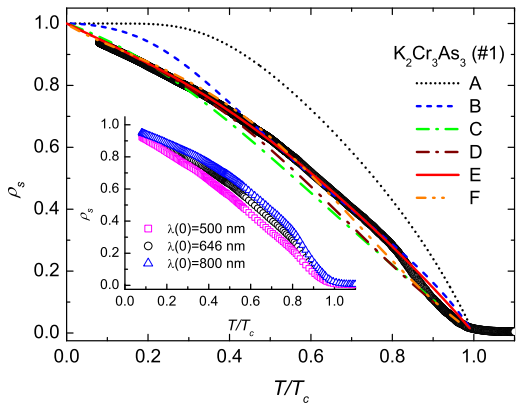}
\end{center}
\caption{Normalized superfluid density $\rho_{\rm s}$ as a function of the reduced temperature $T/T_{\rm c}$, with $\lambda(0)~=~646~$nm. The lines show the calculated  $\rho_{\rm s}$ for different models described in Table.~\ref{ResTab}. The inset shows the superfluid density for changes in $\lambda_0$ of $\sim20\%$.}
\label{Rhos}
\end{figure}

In order to compare the data to models of different gap structures, we converted the penetration depth data to the normalized superfluid density using $\rho_{\rm s}=[\lambda(0)/\lambda(T)]^2$. The value of $\lambda(0)=646~$nm was estimated by $\lambda(0)=\sqrt{\Phi_0H_{c2}(0)}/\sqrt{24\gamma}\Delta_0$, \citep{Lam0Calc} using $H_{c2}(0)=31.2$~T and $\gamma=73$~mJ/mol~K$^2$ from Ref.~\onlinecite{K2Cr3As3Crys} (similar values were also obtained in Ref.~\onlinecite{K2Cr3As3Rep}), as well as a BCS gap magnitude of $\Delta_0=1.76T_c$ (using $k_{\rm B}=1$). Given a gap function $\Delta_k$, the superfluid density $\rho_{\rm s}$ can be calculated by:
\begin{equation}
\rho_{\rm s} = 1 + 2 \left\langle\int_{\Delta_k}^{\infty}\frac{E{\rm d}E}{\sqrt{E^2-\Delta_k^2}}\frac{\partial f}{\partial E}\right\rangle_{\rm FS},
\label{RhoSEq}
\end{equation}

\noindent where $\left\langle\ldots\right\rangle_{\rm FS}$ represents an average over the Fermi surface and $f(E, T)=[1+{\rm exp}(E/T)]^{-1}$ is the Fermi function. Since our measurements were performed on polycrystalline samples, a powder average was calculated following the methods given in Refs.~\onlinecite{pave1} and \onlinecite{pave}. The gap function is given by $\Delta_k(T)=\Delta_0(T)g_k$, where $\Delta_0$ is the maximum gap value and  $g_k$ determines the angular dependence of the gap, which is given for the different models in Table.~\ref{ResTab}. The temperature dependence of the gap is approximated by: \citep{pendepthrev}
\begin{equation}
\Delta_0(T) = \Delta_0(0){\rm tanh}\left[\frac{\pi T_{\rm c}}{\Delta_0(0)}\sqrt{\frac{2}{3}\frac{\Delta C}{\gamma T_c}\left(\frac{T_{\rm c}}{T}- 1\right)}\right],
\label{GapTemp}
\end{equation}

\noindent where $\Delta_0(0)$ is an adjustable parameter, and $\Delta C/\gamma T_c=2.2$ is the specific heat jump at the transition. \citep{K2Cr3As3Crys}

\begin{table}[tb]
\caption{A summary of the different models used to fit the superfluid density in Fig.~\ref{Rhos}. The first column corresponds to the labels in the figure, $g_k$ gives the angular dependence of the gap and $\Delta_0(0)/T_c$ is the gap magnitude in the calculation which best fitted the data. For the pairing state labeled as 2D, Eq.~\ref{RhoSEq} was integrated over a cylindrical Fermi surface, whereas for the remaining states, a spherical Fermi surface was used.}
\label{ResTab}
\begin{ruledtabular}
 \begin{tabular}{c c c c c c }

Label &$g_k$&$\Delta_0(0)/T_c$ & State & Ref.   \\
\hline
\\
A&1 &1.76&  $s$-wave & \\
B& sin$\theta$ &1.6 &  $p$-wave (point) & \citep{pwaveSC} \\
C &sin$\theta$cos$\phi$ &2.5&  $p$-wave (line) & \citep{pwaveSC} \\
D &cos$\theta$ &3.5& $p$-wave (line) & \citep{pwaveSChex} \\
E &cos$2\phi$ &2.2&  2D $d$-wave (line) & \citep{HighTcAnn} \\
F &  \vtop{\hbox{\strut sin($\frac{\sqrt{3}}{2}\pi$ sin$\theta$ sin$\phi$)}\hbox{\strut [cos($\frac{\sqrt{3}}{2}\pi$ sin$\theta$ sin$\phi$)}\hbox{\strut  - cos($\frac{3}{2}\pi$ sin$\theta$ cos$\phi$)]}} &4.5&   $f$-wave (line) & \citep{TheorYiZhou} \\

\end{tabular}
\end{ruledtabular}
\end{table}

To gain insight into the possible pairing states of K$_2$Cr$_3$As$_3$, the experimental data are fitted by using different functional forms of $g_k$, as displayed in Fig.~\ref{Rhos}, where the superfluid density has been compared to several different models. The data are incompatible with the isotropic BCS model (A), since in fully gapped models, $\rho_{\rm s}$ saturates at low temperatures. The data shows no evidence of saturation and $\rho_{\rm s}$ continues to increase with decreasing temperature. Similarly there is a poor agreement with the data for a $p$-wave model with point nodes (B). This corresponds to the A-phase of $^3$He and there are point nodes at the two poles for a three-dimensional Fermi surface. The linear behavior of $\rho_{\rm s}$ at low temperatures is in clear disagreement with this gap structure. The superfluid densities are also displayed for four models with line nodes, two $p$-wave models (C and D), a $d$-wave model (E) and an $f$-wave model (F). Model C has a line node in the $ab$-plane and the size of the gap also reduces to zero with increasing polar angle. Model D has a nodal plane at $k_z=0$, which is similar to the $p_z$ pairing state predicted for small $J/U$, although the pairing is expected to occur predominantly on one of the q1D sheets where there are no nodes. \cite{TheorYiZhou,Theor3}  Model E corresponds to the two-dimensional $d$-wave state often applied to the cuprate superconductors, while the $f$-wave model (F) was proposed as a possible pairing state at large $J/U$ on the three-dimensional band of K$_2$Cr$_3$As$_3$ in Ref~\onlinecite{TheorYiZhou} and has line nodes in the $ab$-plane. It can be seen that models D-F can reasonably account for the temperature dependence of $\rho_{\rm s}$, while one of the $p$-wave models (C) only becomes linear at very low temperatures. The best fits at low temperatures give relatively large values of $\Delta_0(0)/T_c$ for models D and F, indicating strong-coupling superconductivity, as also inferred from the large jump in the specific heat. \cite{K2Cr3As3Rep, K2Cr3As3Crys} The value of the energy gap can be estimated in analogy with the $\alpha$-model of superconductivity, \cite{alphamod} using  $\Delta_0(0)=\Delta_{wc0}(0)\sqrt{(\Delta C/\gamma T_c)\left\langle g_k^4 \right\rangle/1.426\left\langle g_k^2 \right\rangle^2}$, where $\Delta_{wc0}(0)/T_c$ is the weak coupling gap size, taken to be the $d$-wave value of 2.14. \cite{pendepthrev} This expression gives estimates for $\Delta_{0}(0)/T_c$ of 3.6 for models C, D and F, and 3.3 for model E. Therefore the large gap values for models D and F are consistent with the specific heat jump. The gap for model E of 2.2 is near the $d$-wave weak coupling value, while the specific heat jump is considerably larger than the weak coupling value of 0.95. However, we have employed a single band model of the superfluid density with a simplified Fermi surface, therefore self consistent calculations based on the band structure are required in order to precisely fit the data and discriminate between different gap anisotropies. Due to uncertainties in the estimates of the calibration constant $G$ and $\lambda_0$, in the inset of Fig.~\ref{Rhos}, $\rho_{\rm s}$ is plotted for changes in $\lambda_0$ of $\sim20\%$. It can be seen that similar behavior is observed upon adjusting these parameters, although this may also change the best fit values of $\Delta_0(0)/T_c$. In particular, the linear temperature dependence of $\Delta f(T)$ at low temperatures guarantees a linear decrease of  $\rho_{\rm s}$, regardless of the values of  $G$ or $\lambda_0$.

Another possibility is that the line nodes are not imposed by the symmetry of the pairing state but are accidental and arise due to there being a mixture of singlet and triplet states, as a result of inversion symmetry breaking. In this scenario, two gaps open on the split Fermi surface sheets and whether nodes arise on one of the gaps depends on the relative magnitude of the singlet and triplet components, as well as the structure of the triplet state. Such a scenario has been proposed to explain the penetration depth of the noncentrosymmetric Li$_2$Pt$_3$B, where a linear temperature dependence of $\Delta\lambda(T)$ is also observed at low temperatures, indicating the presence of line nodes.  \citep{Li2Pt3Bnodes} A similar situation may arise in K$_2$Cr$_3$As$_3$, which has a sizeable band splitting due to the ASOC. \citep{K2Cr3As3Elec}

\begin{figure}[t]
\begin{center}
 \includegraphics[width=0.85\columnwidth]{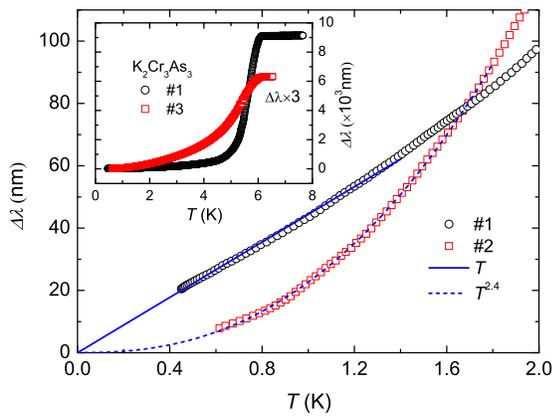}
\end{center}
\caption{Comparison of the low temperature behavior of a high quality sample ($\#1$) and a poor quality sample ($\#3$). The dashed line shows the linear behavior of the high quality sample, while the solid line shows the $T^{2.4}$ dependence of   the poorer sample.  The inset shows $\Delta\lambda$ across the superconducting transition for both samples, where the data for the poor quality sample has been scaled by a factor of three.}
\label{sampqual}
\end{figure}

On the basis of these measurements alone, we can not yet determine the superconducting pairing state and the main result of this study is the linear behavior of $\Delta\lambda(T)$ at low temperatures, which is evidence for the presence of line nodes in the superconducting gap. The linear temperature dependence of $\Delta\lambda(T)$ is reproducible in the best samples with the sharpest superconducting transitions, whereas for poorer quality samples, a different temperature dependence is observed. In Fig.~\ref{sampqual},  $\Delta\lambda(T)$ is displayed for an additional sample (\#3), where the superconducting transition is noticeably broader due to decomposition in the air (see inset), indicating a poorer sample quality. The smaller overall change in $\Delta\lambda$ may also be due to a reduced superconducting fraction. At low temperatures, the behavior is not linear and as shown by the fitted lines, power law behavior of the form $\Delta\lambda(T)\sim T^n$ with $n~=~2.4$ is observed. In general the worse samples display $n>1$, which is expected for systems with line nodes in the presence of impurity scattering. \citep{Scattpen} Although data are shown in  Fig.~\ref{sampqual} for a sample with $n~=~2.4$, the value of $n$ is highly sample dependent with different $n$ being obtained in other poorer quality samples, whereas the linear behavior of the high quality samples can be reproduced and the data points overlap well. This indicates that the linear behavior displayed in Fig.~\ref{LamT} is an intrinsic property of this material. It has previously been suggested that a linear contribution to the superfluid density may also arise in polycrystalline samples due to Josephson coupling between different grains. \cite{JosGran} However, such a scenario does not account for the different behavior in the poor quality samples. Although the effects of decomposition may change the degree of Josephson coupling, this would be expected to change the magnitude of the linear component rather than causing an increase of $n$. In addition, it is found that the temperature dependence hardly changes even when the sample was powdered, \cite{Supp2} indicating that the contribution due to Josephson coupling is small. We also note that preliminary measurements of single crystals also show linear behavior with a similar temperature dependence to polycrystalline samples down to around 0.9~K. However, a small downturn in the frequency shift is observed below 0.9~K,  which is likely due to a small amount of superconducting impurities. These results give further evidence that the linear temperature dependence of the penetration depth cannot be attributed to Josephson coupling between grains in the polycrystalline samples, but is due to the presence of gap nodes. Another indication that the superconductivity of K$_2$Cr$_3$As$_3$ is unconventional comes from the NMR measurements, which show an absence of the Hebel-Slichter coherence peak. \citep{K2Cr3As3NMR} Furthermore, the magnetic field dependence of  $\gamma$ at 2~K of isostructural Rb$_2$Cr$_3$As$_3$ has been reported to display a $\sqrt{H}$ dependence,  \citep{Rb2Cr3As3Rep} which is compatible with nodal superconductivity being found in this study of K$_2$Cr$_3$As$_3$.

To summarize, we have measured the penetration depth of the newly discovered superconductor K$_2$Cr$_3$As$_3$, which has a q1D crystal structure. At low temperatures, distinctly non-BCS-like behavior is observed and the penetration depth $\Delta\lambda(T)$ shows a linear temperature dependence for $T\ll T_c$. Since $\lambda(T)$ is mainly sensitive to low energy excitations, our measurements primarily indicate that K$_2$Cr$_3$As$_3$ is an unconventional superconductor, with evidence for line nodes in the energy gap. These findings will help constrain potential theoretical models of the
pairing state in this compound and suggest that $A_2$Cr$_3$As$_3$ are a new family of unconventional superconductors. Further measurements of high quality single crystals using a range of techniques are desirable to clarify the pairing symmetry of this system.

\begin{acknowledgments}
We thank M.~B.~Salamon, D.~C. Johnston, F.~C.~Zhang, D. F. Agterberg, C.~Cao, Y.~Zhou, J. L. Luo, Y. G. Shi, S.~Kirchner, Z.~A.~Xu, J.~P.~Hu,  X.~Lu and H.~Lee for interesting discussions and their valuable suggestions. This work was supported by the National Basic Research Program of China (No 2011CBA00103), the National Nature Science Foundation of China (No.11474251 and No.11174245) and the Fundamental Research Funds for the Central Universities.
\end{acknowledgments}


\end{document}